\documentclass[aps,pre,twocolumn,groupedaddress,superscriptaddress,showpacs]{revtex4-1}

\usepackage{amsmath}
\usepackage{graphicx}
\usepackage{float}
\usepackage{color}
\usepackage[normalem]{ulem}

\begin{document}

\title{Dynamical design of spatial patterns of colloidal suspensions}

  \author{N. A. M. Ara\'ujo}
   \email{nmaraujo@fc.ul.pt}
   \affiliation{Departamento de F\'{\i}sica, Faculdade de Ci\^{e}ncias, Universidade de Lisboa, P-1749-016 Lisboa, Portugal, and Centro de F\'isica Te\'orica e Computacional, Universidade de Lisboa, Avenida Professor Gama Pinto 2, P-1649-003 Lisboa, Portugal}
   
   \author{D. A. Zezyulin}
   \email{dzezyulin@fc.ul.pt}
   \affiliation{ITMO University, St. Petersburg 197101, Russia}
   \affiliation{Institute of Mathematics with Computer Center, Ufa Scientific Center, Russian Academy of Sciences, Chernyshevskii str., 112, Ufa  450008, Russia}
   \affiliation{Departamento de F\'{\i}sica, Faculdade de Ci\^{e}ncias, Universidade de Lisboa, P-1749-016 Lisboa, Portugal, and Centro de F\'isica Te\'orica e Computacional, Universidade de Lisboa, Avenida Professor Gama Pinto 2, P-1649-003 Lisboa, Portugal}

   \author{V. V. Konotop}
   \email{vvkonotop@fc.ul.pt}
   \affiliation{Departamento de F\'{\i}sica, Faculdade de Ci\^{e}ncias, Universidade de Lisboa, P-1749-016 Lisboa, Portugal, and Centro de F\'isica Te\'orica e Computacional, Universidade de Lisboa, Avenida Professor Gama Pinto 2, P-1649-003 Lisboa, Portugal}

  \author{M. M. Telo da Gama}
   \email{mmgama@fc.ul.pt}
    \affiliation{Departamento de F\'{\i}sica, Faculdade de Ci\^{e}ncias, Universidade de Lisboa, P-1749-016 Lisboa, Portugal, and Centro de F\'isica Te\'orica e Computacional, Universidade de Lisboa, Avenida Professor Gama Pinto 2, P-1649-003 Lisboa, Portugal}

\begin{abstract}
We study the collective dynamics of colloidal suspensions in the presence of a
time-dependent potential, by means of dynamical density functional theory. We
consider a non-linear diffusion equation for the density and show that
spatial patterns emerge from a sinusoidal external potential with a
time-dependent wavelength. These patterns are characterized by a sinusoidal
density with the average wavelength and a Bessel-function envelope with an
induced wavelength that depends only on the amplitude of the temporal oscillations.
As a generalization of this result, we propose a design strategy to obtain a family of
spatial patterns using time-dependent potentials of practically arbitrary shape.
\end{abstract}

\maketitle

\section{Introduction}
The emergence of spatio-temporal patterns in systems driven away from
equilibrium has been always intriguing for scientists and laymen
alike~\cite{Cross1993,Walgraef1997}. Understanding how collective patterns
emerge from the local interactions is not only a question of scientific
curiosity, but also of technological interest, for the resulting patterns are
related in a non-trivial way to the physical properties of the system.

Colloidal suspensions are an example of such systems. With a typical size of
the order of the micron, individual colloidal particles are characterized by
Brownian motion, in very dilute suspensions. However, as the density increases,
correlations among these random walkers lead to their spontaneous
self-organization into mesoscopic structures that extend over length scales
that are much larger than the typical range of the particle-particle
interactions~\cite{Blaaderen1995,Blaaderen2006,Kim2011,Lash15}. 

Several strategies have been explored to control self-organization to drive it
towards desired structures including, for example, the fine tuning of the particle-particle
interactions~\cite{Damasceno12,Sacanna2013,Dias2013}, the control of the
suspending medium~\cite{Silvestre2014}, and the presence of external
constraints such as: interfaces~\cite{Garbin13,Joshi2016},
substrates~\cite{Cadilhe07,Araujo08,Dias2013,Araujo15,Araujo2017}, or
electromagnetic
fields~\cite{Trau1994,Park2011,Furst2014,Klapp2015PRE,LowenRev,Dobnikar,Davies14,Keim2013,Demirors2013,Egelhaaf,Randall2009,Delfau2016}.
In this paper, we will consider the case of electromagnetic fields. They not
only help to control the rotational degrees of freedom~\cite{Davies14} or fine
tune the particle-particle interactions~\cite{LowenRev,Dobnikar,Keim2013}, but
they can also act as virtual molds to induce spatial periodic
patterns~\cite{Demirors2013,Egelhaaf,Randall2009}, as we discuss below.

The time evolution of the system will be described using dynamic
density functional theory~\cite{Tarazona1999,Marconi2000}. Assuming an adiabatic
evolution, it is possible to write down a non-linear diffusion equation for
the local density, from the equilibrium Helmholtz free energy functional. Such
a coarse-grained description still grasps several relevant non-equilibrium
properties and allows accessing the relevant time and length scales of
pattern formation~\cite{Rauscher2010,Goddard2012}.

The paper is organized in the following way. The model and the methods are
discussed in the next section. Results are presented and discussed in
Sec.~\ref{sec:results}. Final remarks are made in Sec.~\ref{sec:conclusions}.

\section{Model and methods}\label{sec:model}
Let us consider a system of colloidal particles in the overdamped regime. The
interaction between two particles is described by a purely-repulsive pairwise
potential $V(r)$, where $r$ is the distance between
particle centers. The potential associated to the external field
$V_\mathrm{ext}$ is assumed periodic (sine function) along the $x$-direction.
For stationary periodic potentials, particles tend to accumulate along the
minima of the potential forming bands along the $y$-direction~\cite{Nunes2016}.
In order to study the effect of time-dependent potentials, we consider that the
characteristic wavelength of the external field creating the spatial lattice oscillates
periodically in time around unity. Thus the particle-field interaction is
described by the potential,
\begin{equation}\label{potential}
V_\mathrm{ext}(\vec{r},t)=V_0\sin{\left(\left[1-V_1\sin\left(\omega
t\right)\right]x\right)} \ \ ,
\end{equation}
where $\omega$ is a frequency and $t$ is the time in units of the Brownian time
$\tau_B=r_p^2\gamma(k_BT)^{-1}$ (the time over which a colloidal particle
diffuses over a region equivalent to $r_p^2$), where $r_p$ is the radius of the
colloidal particles, $k_B$ is the Boltzmann constant, $T$ is the temperature,
and $\gamma$ the Stokes coefficient. $V_0$ and $V_1$ are the amplitudes of the
potential and oscillations, respectively. In the limit $V_1=0$, the stationary
potential with unit wavelength is recovered, setting the units of length,
while for $V_0=0$, purely-repulsive particles are expected to be distributed
uniformly in space.

\begin{figure*}[t]
  \begin{centering}
    \includegraphics[width=\textwidth]{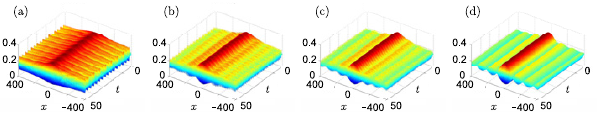}
\end{centering}
\caption{Time evolution of the density profile $\rho(x,t)$, for a 1D system
	with $V_1=0.025$, and different values of $\omega$, namely: (a) $1$,
	(b) $2$, (c) $3$, and (d) $10$. The results were obtained by
	solving Eq.~(\ref{eq::diffusion.eq}) numerically, in a domain with size $800$ and periodic boundary conditions. Asymptotically, a
spatially varying pattern emerges with a characteristic wavelength that
decreases with $V_1$.~\label{fig:3d}}
\end{figure*}

The Helmholtz free energy functional of the system can be approximated
as~\cite{Tarazona1999},
\begin{equation}
\begin{split}
	&\mathcal{F}[\rho(\vec{r},t)]=k_{B}T\int
	\rho(\vec{r},t)\left[\log\left(\rho(\vec{r},t)\Lambda^{2}\right)-1\right]d\vec{r}
	\\ &+\frac{1}{2}\int\int \rho(\vec{r},t) \rho(\vec{r}',t)
	V(\left|\vec{r}-\vec{r}'\right|)d\vec{r}'d\vec{r} \\
	&+\int\rho(\vec{r},t)V_{ext}(\vec{r},t)d\vec{r} \ \ ,
	\label{functional}
\end{split} 
\end{equation}
where, $\Lambda$ is the thermal de Broglie wavelength (with units of length) and
$\rho(\vec{r},t)$ is the (number) density, defined as the number of colloidal
particles per unit area, in two dimensions. The first term is the free energy
of the ideal gas, the second term corresponds to a mean-field approximation of
the pairwise correlations, and the third one is the interaction with the
external field.

From the dynamic density-functional theory (DDFT), assuming adiabatic evolution
of the system, one can obtain the time evolution of the density from the
equilibrium Helmholtz free energy functional~\cite{Tarazona1999},
\begin{equation}\label{eq:density.evol}
  \gamma\frac{\partial\rho(\vec{r},t)}{\partial t}=
  \nabla\left[\rho(\vec{r},t)\nabla
  \frac{\delta\mathcal{F}[\rho(\vec{r},t)]}{\delta\rho(\vec{r},t)}\right].
\end{equation}
Assuming the local density approximation (LDA)~\cite{Nunes2016}, 
$\rho(\vec{r}',t)\approx
\rho(\vec{r},t)+\left(\vec{r}'-\vec{r}\right)\nabla\rho$, and the functional
defined in Eq.~(\ref{functional}), one can obtain a non-linear diffusion
equation for the time evolution of the density,
\begin{equation}
	\gamma\frac{\partial\rho}{\partial
	t}=\nabla\left[A\rho\nabla\rho+\frac{\partial V_\mathrm{ext}}{\partial
	x}\rho \mathbf{i}\right]+k_BT\Delta\rho \ \ . \label{eq::diffusion.eq}
\end{equation}
Here, $\mathbf{i}$ is the unit vector directed along the $x$-axis, and $A=\int
V\left(\vec{r}'\right)d\vec{r}'$, where the integral is over the entire space,
is a positive constant for a purely-repulsive pairwise
interaction~\cite{Nunes2016}. The first two terms are related to the
particle-particle and particle-field interactions, while the third one
results from the interaction with the suspending medium.

\section{Results}\label{sec:results}
Given the translational invariance of $V_\mathrm{ext}(\vec{r},t)$ along the
$y$-direction (see Eq.~(\ref{potential})), we solve numerically a 1D version of
Eq.~(\ref{eq::diffusion.eq}) in a finite domain, with periodic boundary
conditions, to reduce finite-size effects. The domain size is such that
$x\in\left[-400,400\right]$ in length units; numerical results for a smaller
domain $x\in\left[-200,200\right]$ revealed no significant dependence on
the domain size. Without loss of generality, we consider $\gamma=1$ and
$k_BT=1$, which can be mapped to any other values by properly rescaling $x$ and
$t$, in Eq.~(\ref{eq::diffusion.eq}). $k_BT=1$ sets the energy scale, such that
$V_0$ is defined in units of $k_BT$.  Rescaling the mean density, one can also
assume that $A\in\{\pm 1,0\}$. As initial conditions, we considered a uniform
density profile ($\rho_0=1/(2\pi)$). Tests with different initial conditions,
but the same average density, revealed no dependence of the final pattern on
the initial conditions.

As summarized in Fig.~\ref{fig:3d}, spatio-temporal patterns are obtained in
the presence of the time-dependent field. While for slow variations of the
wavelength of the potential (low $\omega$), the density follows the time evolution
of the potential (Figs.~\ref{fig:3d}(a)-(c)), for sufficiently rapid variations
(high $\omega$), a spatial pattern emerges with a stationary envelope and a
non-stationary filling, that persists over time (Fig.~\ref{fig:3d}(d)).
However, the characteristic wavelength of the emerging pattern does not seem to
depend on the value of $\omega$. We also observe dynamical localization at the
center of the box, which is not observed in the stationary case ($V_1=0$). This
is a consequence of the shape of the potential~(\ref{potential}), since 
at $x=0$ the value of the potential does not change with time.

\subsection{Effect of the averaged potential}
\begin{figure}[t]
\begin{centering}
\includegraphics[width=\columnwidth]{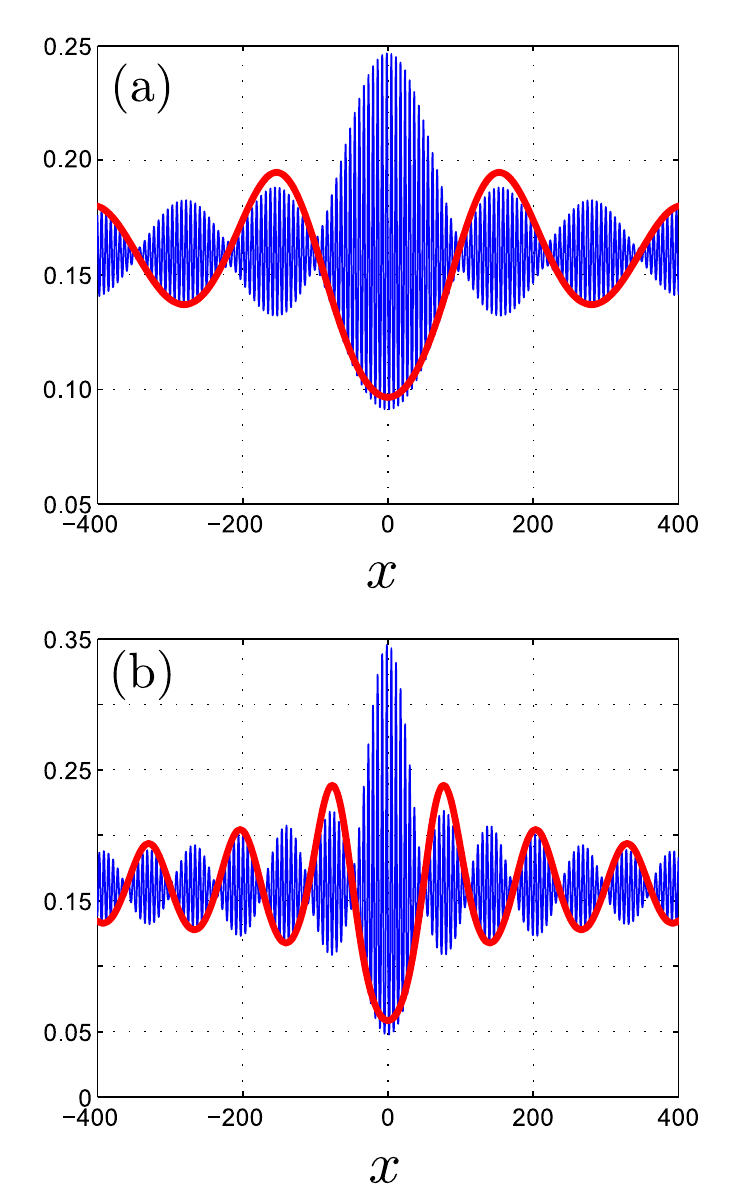}\\
\end{centering}
\caption{Density, $\rho(x)$, at $t=0$, for a 1D system. The blue curve
	was obtained by solving Eq.~(\ref{eq::diffusion.eq}), in a
	domain with size $800$ unit lengths and periodic boundary conditions,
	for $A=0$, (a) $V_0=0.5$, $V_1=0.025$ and (b) $V_0=1$, $V_1=0.05$. The
	red curves are given by $B\exp\left[-V_0J_0(V_1x)\right]$, where $B$
	was adjusted to fit the numerical data.\label{fig:bessel}}
\end{figure}
The appearance of a new spatial scale in a time-varying potential can be
understood using averaging arguments. At high enough $\omega$, we
assume that the temporal dependence of the density is much slower than the
variation of the potential. Then, the density can be considered approximately
constant during one period of the potential oscillation, i.e., in the
interval $(t, t+2\pi/\omega)$, $\omega\gg1$. Performing the average over
one period (i.e., multiplying by $2\pi/\omega$ and integrating from  $t=\tau$
to $t=\tau+2\pi/\omega$), we conclude that the rapidly oscillating
potential can be approximated by an effective averaged potential:
\begin{equation}
\langle V (x) \rangle = \frac{\omega}{2\pi}\int_0^{2\pi/\omega}
V_{ext}(x,t)dt.
\end{equation}

Thus, instead of Eq.~(\ref{eq::diffusion.eq}), we consider 
\begin{equation}
\frac{\partial\rho}{\partial t}=\nabla\left[A\rho\nabla\rho+\frac{d \langle V
(x) \rangle}{d x}\rho \mathbf{i}\right]+ \Delta\rho,
\end{equation}
where, according to the above discussion, we rescaled $\gamma$ and $k_BT$ to
unity. To obtain a time-independent solution, we solve the 1D version
of this equation, 
\begin{equation}
\left[A\rho_x\rho+ { \langle V (x) \rangle}_x\rho\right]_x+ \rho_{xx}=0 ,
\end{equation}
where $\rho_x$ and $\rho_{xx}$ are the first and second spatial derivatives
along the $x$-direction, respectively.  This equation can be integrated, which
gives $A\rho_x\rho+ {\langle V(x) \rangle_x}\rho + \rho_x=C_0$, where $C_0$ is
some constant. We set $C_0=0$, corresponding to $\rho_x=0$ when $\rho=0$, and we
integrate the resulting equation again, which gives
\begin{equation}\label{eq:average}
A\rho +\ln\rho = -\langle V \rangle + C_1.
\end{equation}

For the particular choice (\ref{potential}), we find
\begin{equation}
\langle V \rangle  = V_0\sin(x)J_0(V_1x),
\end{equation}
where $J_0$ is the zero-order Bessel function. Thus, if $V_1\ll 1$, as in
Fig.~\ref{fig:3d}, the average potential consists of the product of a rapidly
oscillating $\sin(x)$ and a slowly varying envelope  $J_0(V_1x)$.  The latter
can be used to compute the smooth envelope of the density $\rho_s$:
\begin{equation}\label{eq:average_s}
A\rho_s +\ln\rho_s = -V_0 J_0(V_1x) + C_2 .
\end{equation}
When $A=0$ this equation admits a simple analytical solution; for nonzero $A$ it has to be solved numerically. The constant $C_2$ is determined from the value
of the mean density. Thus the slow spatial scale is described by the
oscillating zero-order Bessel function $J_0(V_1x)$. At large $|x|$, the
distance between the successive zeros of $J_0(x)$ approaches $\pi$. Thus the
distance between the ``nodes'' of the pattern $\rho_s$  can be estimated as
$\pi/V_1$, which is in good agreement with the numerical results.

In order  to make a simple check, we consider $A=0$ in
Eq.~(\ref{eq:average_s}) and obtain
\begin{equation}
	\rho_s = B\exp[-V_0J_0(V_1 x)],
\end{equation}
where $B=\exp(C_2)$. Thus, the slow envelope is proportional to
$\exp[-V_0J_0(V_1x)]$, in agreement with the observed pattern,
see Fig.~\ref{fig:bessel}. For nonzero $A$, Eq.~(\ref{eq:average}) is
solved numerically (see Fig.~\ref{fig:cosine}  and the example for $A=1$, in
the next section).

\subsection{Designing spatial patterns}
\begin{figure}
\begin{centering}
\includegraphics[width=\columnwidth]{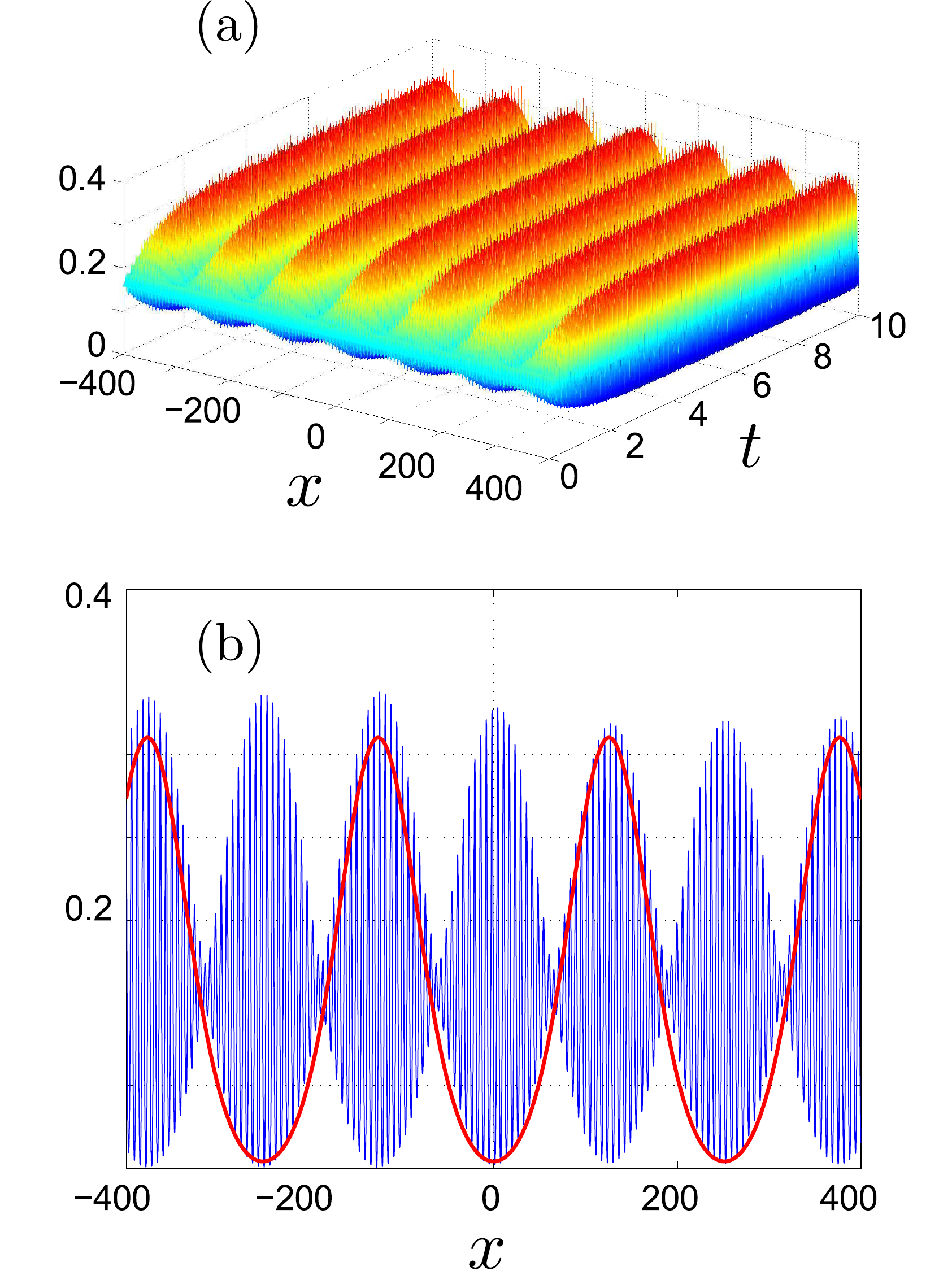}
\end{centering}
\caption{(a) Time evolution of the density profile $\rho(x,t)$, obtained by
	solving Eq.~\ref{eq:average} in a 1D domain with $800$ unit
	lengths and periodic boundary conditions. $V_\mathrm{ext}$
	corresponds to a cosine-shaped potential, obtained from
	Eq.~(\ref{eq:exp_pot}), truncated after eight terms. (b) Comparison of
	the density profile $\rho(x)$, at $t=10$ (blue curve), with the slow
	envelope $\rho_s$ (red curve) found numerically from
	Eq.~(\ref{eq:slow}). The other parameters are $A=1$, $\omega=20$, and
$V_1=0.025$.\label{fig:cosine}}
\end{figure}
In the previous section, we have shown that a spatial pattern emerges from the
interaction with a time-dependent potential. We now develop a framework to
reverse-engineer the pattern and determine the external potential that 
leads to a given pattern. To do so, let us first show how to use a combination
of sine potentials with time-dependent wavelengths to obtain an effective average
potential with a slow cosine-shaped envelope.

Let us consider a potential in the form of the superposition
$V_{ext}=\sum_{n=0}^\infty V_n$, where a $2\pi/\omega$-periodic (in time)
elementary block reads
\begin{equation}\label{eq:potential_series}
\begin{split}
V_n=V_0^{(n)}\sin{\left(\left[1-V_1^{(n)}\sin\left(\omega t\right)\right]x +
2n\omega t\right)}, \\ \quad n=0,1,\ldots .
\end{split}
\end{equation}
Notice that Eq.~(\ref{potential}) was generalized by adding a time-dependent
spatial shift $2n\omega t$, i.e., now in the expansion over the propagation,
non-stationary waves are used. The average of the $n$th elementary block reads 
\begin{equation}
\langle V_n \rangle  = V_0^{(n)}\sin(x)J_{2n}(V_1^{(n)}x),
\end{equation}
where $J_{2n}$ is the $2n$th Bessel function. Thus the average of the
superposition has the form
\begin{equation}\label{eq:exp_pot}
\langle V \rangle  = \sin(x) \sum_{n=0}^\infty V_0^{(n)}J_{2n}(V_1^{(n)}x) \ \
\end{equation}
and the density profile is given by Eq.~(\ref{eq:average}). 

We now consider an example of another average potential that can be
dynamically engineered. To this end, we use the expansion of the cosine
function on Bessel functions,
\begin{equation}\label{eq:cosBes}
\cos z = J_0(z) - 2J_2(z) + 2J_4(z) - 2J_6(z) + \ldots.
\end{equation}
Thus, for $V_1^{(1)}=V_1^{(2)}=\ldots=V_1$, $V_0^{(0)}=1$,
$V_0^{(1)}=V_0^{(3)}=V_0^{(5)}=\ldots =-2$, and
$V_0^{(2)}=V_0^{(4)}=V_0^{(6)}=\ldots =2$, the average potential is
\begin{equation}
\langle V \rangle  = \sin(x)\cos(V_1x).
\end{equation}
For small $V_1$, the average potential has the structure of the slow envelope
$\cos(V_1x)$ and fast filling $\sin(x)$.  Omitting the latter contribution,
using Eq.~(\ref{eq:average}) we obtain for the smooth envelope $\rho_{s}$ of the
density
\begin{equation}\label{eq:slow}
A\rho_s +\ln\rho_s = -V_0\cos(V_1 x) + C_4 ,
\end{equation}
where $C_4$ is a constant. When $A=0$, this equation admits a simple analytical
solution; for nonzero $A$ it should be solved numerically. Recall that the
constant on the right-hand side is obtained from the value of the mean density.
In practice, the expansion~(\ref{eq:cosBes}) is truncated at a finite number of
terms, and therefore the resulting averaged potential will be affected by the
truncation. Figure~\ref{fig:cosine} depicts the numerical results for eight terms
in the expansion~(\ref{eq:cosBes}). The resulting pattern (blue line in
Fig.~\ref{fig:cosine}(b)) is compatible with the envelope $\rho_s$, estimated
from Eq.~(\ref{eq:slow}), corresponding to the red line.

Even more generally, normalizing the interval along the x-axis to
$\left[0,1\right]$, one can use the fact that the Bessel functions form a complete set,
i.e., on a given interval $\left[0,1\right]$, the Bessel functions
$\sqrt{x}J_{n}\left(V_k^{(n)}x\right)$, constitute a complete set, and thus any
functional dependence of the envelope on the average potential ($\langle V
\rangle(x)$) can be expanded in Bessel functions. Thus, with a proper
parameterization of the potential given by Eq.~(\ref{eq:potential_series}), one
can obtain the targeted envelope.

In short, to obtain a given periodic pattern, one needs to invert
Eq.~(\ref{eq:average}) to compute the average potential $\langle V(x)\rangle$,
which supports the given pattern. Then, the average potential is
expanded in terms of Bessel functions (or Fourier series over cosines) and the
potential in Eq.~(\ref{eq:potential_series}) is parameterized accordingly.

\section{Final remarks}\label{sec:conclusions}
We have studied the collective dynamics of colloidal suspensions in the
presence of time-varying potentials. Using dynamic density-functional theory,
the local interactions can be coarse-grained to obtain a non-linear diffusion
equation for the time evolution of the density, allowing us to access the
relevant time and length scales to observe the formation of spatio-temporal
patterns. For a potential with sinusoidal spatial symmetry and a characteristic
wavelength that varies periodically with time, we have shown that a spatial
pattern emerges, at high enough frequencies, with a stationary envelope.
However, the new characteristic length, does not seem to depend on the
frequency of oscillations but rather on their amplitude. Using averaging
arguments, we obtained an equation for the envelope of the pattern and its
dependence on the amplitude of the oscillations. This equation is in good
agreement with the patterns obtained by numerical solution of the non-linear
diffusion equation. The patterns described here resemble Faraday patterns,
which typically appear in vibrating recipients. Different from those patterns,
here it is the wavelength of the external field that oscillates rather than the
magnitude of the field.

After establishing the relationship between the potential and the properties of
the emerging pattern, it was possible to develop a reverse engineering strategy
to tune the pattern by changing the amplitude and frequency of the potential
oscillations. Any pattern envelope that can be expanded over a series of Bessel
functions can be obtained with this method. To access the relevant time and
length scales, we have considered a coarse-grained (continuum) description of
the colloidal suspension, assuming that the size of the colloidal particles is
much smaller than the other length scales. For values of the average wavelength
comparable to the particle size, one expects that the discrete nature of the
particles plays a role, and other methods should be considered. In particular,
it is interesting to check under what conditions the rapidly varying (sine)
term of the density profile becomes irrelevant (a constant) and the profile
given by the slowly varying envelope.

The framework developed here can be extended to other potentials and patterns.
For convenience, we have considered expansions in Bessel functions, but other
complete sets could be used. For example, under certain conditions, the
Fourier-Bessel expansion can be considered instead, where a complete set is
defined using the orthogonal versions of the Bessel function of the first kind.
Future work may consider these possible extensions.

\section{Acknowledgements}
We acknowledge financial support from the Portuguese Foundation for Science and
Technology (FCT) under Contracts nos. UID/FIS/00618/2013 and
EXCL/FIS-NAN/0083/2012. DAZ also acknowledges financial support from Russian
Science Foundation (grant No. 17-11-01004).

\bibliography{paper}

\end{document}